\documentclass[prd,preprint,superscriptaddress,amsmath,amssymb]{revtex4}

 \pdfoutput=1

\usepackage{graphicx}
\usepackage{color,slashed}

\begin{document}

\begin{flushright}{KIAS-P20012}
\end{flushright}

\title{\bf   Electron and muon $g-2$, radiative neutrino mass, and $\ell'\to \ell \gamma$ in a  $U(1)_{e-\mu}$ model}

\author{Chuan-Hung Chen}
\email{physchen@mail.ncku.edu.tw}
\affiliation{Department of Physics, National Cheng-Kung University, Tainan 70101, Taiwan}

\author{Takaaki Nomura}
\email{nomura@kias.re.kr}
\affiliation{School of Physics, KIAS, Seoul 02455, Korea}

\date{\today}

\begin{abstract}

A nonconventional $U(1)_{e-\mu}$ gauge model is proposed to explain the observed  neutrino masses and the unexpected anomalous magnetic  moments of the electron and muon (lepton $g-2$), where for suppressing the neutrino coupling to $Z'$ gauge boson, only the right-handed electron and muon in the standard model carry the $U(1)_{e-\mu}$ charge.  Although the light lepton masses are suppressed when the gauge symmetry is spontaneously broken, they can be generated through the Yukawa couplings to newly introduced particles, such as vector-like lepton doublets and singlets, and scalar singlets. It is found that the same Yukawa couplings combined with the new scalar couplings to the Higgs can induce the radiative lepton-flavor violation processes $\ell' \to \ell \gamma$ and lepton $g-2$, where the lepton $g-2$ is proportional to $m_{\ell}$. When Majorana fermions and a scalar singlet are further added into the model, the active neutrinos can obtain masses via the radiative seesaw mechanism. When the bounds from the $m_e$ and $m_\mu$ and the neutrino data are satisfied, we find that the electron $g-2$ can reach an order of $-10^{-12}$, and the muon $g-2$ can be an order of $10^{-9}$. In addition, when the $\mu\to e \gamma$ decay is suppressed, the resulting branching ratio for $\tau\to e \gamma$ can be of $O(10^{-8})$, and that for $\tau\to \mu \gamma$ can be as large as the current upper limit. 

\end{abstract}

\maketitle

\section{Introduction}

 The observed neutrino oscillations indicate that  neutrinos are massive particles~\cite{PDG}, and combining with the cosmological results~\cite{Couchot:2017pvz}, it is found that the neutrino masses have to be much below eV. Although several resolutions have been proposed to explain the neutrino masses, such as type-I seesaw \cite{SeeSaw}, type-II seesaw \cite{ Magg:1980ut,Konetschny:1977bn}, and radiative seesaw~\cite{Ma:2006km} mechanisms, the real mechanism is not yet concluded.

A potential hint for new physics has been found in  the muon anomalous magnetic moment (muon $g-2$) since the E821 experiment at Brookhaven National Lab (BNL)~\cite{Bennett:2006fi} reported a $3.3\sigma$ deviation from the standard model (SM) prediction, which is shown as~\cite{PDG,Davier:2019can}:
 \begin{equation}
 \delta a_\mu = a^{\rm exp}_\mu - a^{\rm SM}_\mu = (26.1\pm 7.9)\times 10^{-10}\,. \label{eq:da_mu}
 \end{equation}
A $3.7\sigma$ deviation was also obtained by the lattice calculations as $\delta a_\mu = (27.4\pm 7.3)\times 10^{-10}$~\cite{Blum:2018mom} and $\delta a_\mu = (27.06\pm 7.26)\times 10^{-10}$~\cite{Keshavarzi:2018mgv}.  Due to the discrepancy between the experimental measurement and the theoretical prediction, various solutions have been proposed to resolve the anomaly over the years~\cite{Czarnecki:2001pv,Gninenko:2001hx,Ma:2001mr,Chen:2001kn,Ma:2001md,Padley:2015uma,Benbrik:2015evd,Nomura:2016rjf,Baek:2016kud,Altmannshofer:2016oaq,Chen:2016dip,Lee:2017ekw,Chen:2017hir,Das:2017ski,Calibbi:2018rzv,Barman:2018jhz,Kowalska:2017iqv,Chen:2019nud,Nomura:2019btk,Chen:2019okl,Cao:2019evo,Chen:2020ptg,Kumar:2020web,deJesus:2020ngn,Kawamura:2020qxo,Iguro:2020rby,Han:2020exx}. Although the recent result on the hadron vacuum polarization (HVP), which was calculated by  Budapest-
Marseille-Wuppertal (BMW) collaboration~\cite{Borsanyi:2020mff}, weakens the necessity of a new physics effect, it is shown in~\cite{Crivellin:2020zul,Keshavarzi:2020bfy,Passera:2008jk} that the BMW result leads to new intensions with the HVP extracted from $e^+ e^-$ data and the global fits to the electroweak precision observables.  

The new muon $g-2$ measurements performed in the E989 experiment at Fermilab and the E34 experiment at J-PARC aim for a precision of 0.14 ppm~\cite{Grange:2015fou} and 0.10 ppm~\cite{Otani:2015jra},  in which   the experimental accuracy can be improved by a factor of $4$ and $5$, respectively. If we assume the future experimental and theoretical uncertainties can be respectively reduced by a factor of 4 and $\sqrt{2}$,  it is expected that with a 3$\sigma$ measurement, $\delta a_\mu \approx (12 \pm 4)\times 10^{-10}$ can be observed by the Fermilab muon $g-2$ experiment~\cite{Grange:2015fou}.

Applying  the most accurate measurement of the fine structure constant, which is measured using $^{133}$Cs,  to the theoretical calculations~\cite{Aoyama:2014sxa,ATN2019}, it is found that the difference in the electron $g-2$ between the experiment and the SM result has a $2.4\sigma$ deviation and is expressed as~\cite{PYZEM}:
\begin{equation}
\delta a_e = -(8.8\pm 3.6)\times 10^{-13}\,.  \label{eq:da_e}
\end{equation}
 Differing from the muon $g-2$, the electron $g-2$ experimental value is lower than the SM result. In order to simultaneously  explain the anomalistic electron and muon $g-2$, some possible resolutions are provided in studies  in the literature~\cite{Giudice:2012ms,Aboubrahim:2014hya,Abada:2014nwa,Aboubrahim:2016xuz,Marciano:2016yhf,Davoudiasl:2018fbb,Crivellin:2018qmi,Liu:2018xkx,Han:2018znu,Endo:2019bcj,Abdullah:2019ofw,Gardner:2019mcl,Badziak:2019gaf,CarcamoHernandez:2019ydc,Hiller:2019mou,Haba:2020gkr,Bigaran:2020jil,Calibbi:2020emz,Jana:2020pxx}.
  
 In order to simultaneously explain  the observed neutrino masses and  the lepton $g-2$ anomalies, we investigate a nonconventional gauged  $U(1)_{e-\mu}$ extension of the SM.    Due to the  $U(1)_{e - \mu}$ symmetry, the mass terms of the active neutrinos and introduced Majorana fermions are suppressed; thus, no neutrino mass is generated at the tree level.  As a result,  all relevant phenomena are induced from the one-loop effects. It is known that  a light $Z'$ using for explaining the muon $g-2$ is excluded by BaBar~\cite{Lees:2017lec} and NA64~\cite{Banerjee:2016tad} experiments though the invisible decays.  Moreover, the potential strict constraints for a light $Z'$ are from the $\nu-e$ scattering ~\cite{Harnik:2012ni}  and the  neutrino trident production experiments~\cite{Altmannshofer:2014pba,Altmannshofer:2019zhy}.  In order to escape from the neutrino-related experimental constraints,  unlike  the conventional $U(1)_{e-\mu}$, where the associated $Z'$-gauge boson couples to the right-handed lepton singlets and the left-handed lepton doublets in the SM~\cite{He:1990pn}, we propose that the $U(1)_{e-\mu}$ model only couples to the right-handed leptons. 

 The immediate problem with the $U(1)_{e-\mu}$  model is the massless electron and muon.  To resolve this problem, we add new representations into the model, such as vector-like lepton doublets and singlets, and scalar singlets, where with the exception of one vector-like lepton singlet,  they all carry the $U(1)_{e-\mu}$ charges. Thus, the lepton masses can be generated through the mixings with the introduced heavy charged leptons at the tree level. 
 
 It is found that when the new scalar couplings are considered, the same effects, which lead to the light lepton masses, can induce the radiative lepton-flavor violation (LFV) processes at the one-loop level. Taking the initial and final leptons to be the same species, the electron and muon $g-2$ can then be generated. Because the effect on the $\tau$ $g-2$  is small, we do not further discuss the influence on the $\tau$-lepton. We note that  since the $Z'$-gauge boson only couples to the right-handed light leptons,  the induced lepton $g-2$ values are negative~\cite{Jegerlehner:2009ry, Lindner:2016bgg}, and   the resulting ratio  is $\delta a^{\rm Z'}_e /\delta a^{\rm Z'}_\mu \sim m^2_e/m^2_\mu$.  If we use the $Z'$ effect as the single source leading to  the negative electron $g-2$, the resulting muon $g-2$ is also negative and contradicts the indications in the current data. Therefore, in this study, the $g_{Z'}$ gauge coupling and $m_{Z'}$ have to be taken in such a way that the induced muon $g-2$ is small enough.  Hence, the main source for the lepton $g-2$ anomalies is not from $Z'$ effects but from those introduced for obtaining the electron and muon masses.  We will show that the observed lepton $g-2$ anomalies and the light lepton masses can be accommodated in the model.

We further find that when two Majorana fermions and one scalar singlet, which carry the $U(1)_{e-\mu}$ charge, are introduced, the neutrino mass can be radiatively produced through the one-loop Feynman diagrams.  Since some of the involved parameters are related to parameters  that  contribute to $m_{e,\mu}$ and $\delta a_{e,\mu}$, it is found that when the bounds from the current neutrino data are satisfied, besides the fact that $m_e$ and $m_\mu$ can fit the experimental values, the results of  $\delta a_e\sim O(-10^{-12})$ and $\delta a_\mu \sim O(10^{-9})$ can be  obtained. 

When the rare $\mu\to e \gamma$ decay is suppressed in the study, and all the relevant constraints are satisfied, we find that in the model,  the branching ratio (BR) for $\tau \to e \gamma$ can be under the current experimental upper bound. When we use constrained parameter values to estimate the BR for $\tau\to \mu \gamma$, it is found that $BR(\tau\to \mu \gamma)$ can be over the current upper limit; that is, the $\tau\to \mu \gamma$ decay can be used to further constrain the free parameter space. Nevertheless,  the results of $\delta a_e \sim O(-10^{-12})$ and $\delta a_\mu \sim O(10^{-9})$ are not influenced. 

The paper is organized as follows: We  introduce the model and discuss the relevant Yukawa couplings and the scalar potential in Sec. II. The vacuum stability conditions are also analyzed in this section. We discuss the tree-level charged lepton mass matrix and the loop-level  neutrino mass matrix in Sec. III. In Sec. IV, we formulate the radiative LFV processes and lepton $g-2$, and the numerical analysis is shown in Sec. V. We provide a summary in Sec. VI.

\section{ Model}

In order to explain the neutrino data and the electron and muon $g-2$, we consider an anomaly-free gauged $U(1)_{e-\mu}$ extension of the SM~\cite{He:1990pn}, where the associated $Z'$-gauge boson only couples to the right-handed electron and muon. We  add new representations, such as two vector-like lepton doublets $(X^\ell, \ell=e,\mu)$, one vector-like lepton singlets ($X$), two right-handed  neutrino singlets $(N^\ell)$,   three  scalar singlets $S^\ell, S$.  In addition, for the $U(1)_{e-\mu}$ gauge anomaly cancellation, we need to introduce two vector-like lepton singlets ($\Sigma^\ell$), where their masses arise from the new scalar singlet $\eta$. For these fields, we impose a $Z_3$ symmetry to suppress the interactions with the SM fermions.  
The representations and charge assignments of particles are given in Table~\ref{tab:reps}. 
The other SM particles, which are not shown in the table,  do not have the $U(1)_{e-\mu}$ charges. 

\begin{table}[htp]
\caption{Representations and charge assignments of particles in $SU(2)_L\times U(1)_Y \times U(1)_{e-\mu} \times Z_3$ where $\omega^3 =1$ with $\omega^* = \omega^2$.}
\begin{center}
\begin{tabular}{c|ccccccccc}
  & ~$e_R/\mu_R$ ~& ~$X^{e/\mu}_{L,R}$ ~&  ~$N^{e/\mu}$ ~& ~$S^{e/\mu}$~ & ~$S$~ & ~$X_{L,R}$~ & ~$\Sigma_{L(R)}^e$~ & $\Sigma_{L(R)}^\mu$  & $\eta$  \\ \hline
  $SU(2)_L$ & 1 & 2 & 1 & 1 & 1 & 1&1 & 1 & 1\\ \hline 
  $U(1)_Y$ & $-1$ & $-1/2$ & $0$ & $0$ & $0$ & $-1$ & $-1$  & $-1$ & 0\\ \hline
  $U(1)_{e-\mu}$ & $1/-1$ & $1/-1$ & $1/-1$ & $1/-1$ &  $2$ & $0$ & $1(0)$ & $-1(0)$ & $1$ \\ \hline 
  $Z_3$    & 0 & 0 & 0 & 0 &  0 & $0$ & $\omega^2 (\omega)$ & $\omega(\omega^2)$ & $\omega$ \\ \hline
\end{tabular}
\end{center}
\label{tab:reps}
\end{table}%

  %

 The gauge invariant Yukawa couplings for the lepton sector can be partly written as:
 \begin{align}
 -{\cal L}_Y & = \bar L_\tau y_\tau H \tau_R +\bar X_L y_\ell \ell_R S^{\ell \dagger} +\bar L_{\ell'} \tilde{y}_{ \ell'} H X_R + \bar L_{\ell'} y^\ell_{\ell'} X^\ell_R S^{\ell \dagger} \nonumber \\
 & + \overline{X^\ell_L}  y^\ell_X H \ell_R +\overline{X^\ell_L} \tilde{y}^\ell_X \tilde{H} N^\ell + h_e N^{eT} C N^e S + h_\mu N^{\mu T} C N^\mu S^\dagger\nonumber \\
  &  +M_{X^\ell} \overline{X^\ell_L} X^\ell_R + M_X \bar X_L X_R + m_{N_{\mu e}} N^{eT} C N^\mu+ H.c.\,, \label{eq:Yukawa}
 \end{align}
 where  $C= i \gamma^0 \gamma^2$, and $L$ and $H$ are respectively the SM lepton and  Higgs doublets;  $\ell=e, \mu$, and  $\ell'$ denotes all of the SM lepton-flavor indices.  Since $\tau_R$ does not carry the $U(1)_{e-\mu}$ charge, after electroweak symmetry breaking (EWSB), the tau-lepton can obtain mass through the Higgs mechanism and its mass is expressed as $m^0_\tau= y_\tau v/\sqrt{2}$, where $v$ is the vacuum expectation value (VEV) of $H$. Although the electron and muon masses are suppressed in Eq.~(\ref{eq:Yukawa}), we will show that their masses can be induced through the mixings with $X^{\ell}$ and $X$. 
 The Yukawa couplings related to $\Sigma^\ell$ are expressed as:
 \begin{align}
 -{\cal L}_{\Sigma^\ell} = &  y_\Sigma^e \bar \Sigma^e_L \Sigma^e_R \eta  + y^\mu_\Sigma \bar \Sigma^\mu_L \Sigma^\mu_R \eta^\dagger + \overline{\Sigma^e}_L \Sigma^\mu_R ( f_e S^e + f_\mu S^{\mu \dagger} ) \nonumber \\
 & + \overline{\Sigma^\mu_L} \Sigma^e_R ( g _e S^{e\dagger} + g_\mu S^\mu) +H.c.  \label{eq:sigma}
 \end{align}
In the model, the VEVs of $S^\ell$ will be taken to be around $1$ GeV; thus, the $\Sigma^\ell$ masses are mainly dictated by the VEV of $\eta$ and are formulated as $m_{\Sigma^\ell} \approx  y^\ell_{\Sigma} v_\eta/\sqrt{2}$. Since the role of $\Sigma^\ell$ is used to cancel the $U(1)_{e-\mu}$ gauge anomaly, their effects are irrelevant to the study. Therefore, we will not further discuss the effects in Eq.~(\ref{eq:sigma}) in the following analysis. 

Since the electron and muon masses, the lepton $g-2$, and the neutrino masses are strongly correlated to the VEVs of scalar fields and the scalar couplings in the scalar potential, we have to discuss the vacuum stabilities of  the scalar fields.   We note that the singlet scalar $\eta$ is introduced to obtain the $\Sigma^\ell$ masses. Although it can couple to other scalar fields, because these couplings do not significantly affect the phenomena, which we study; for simplicity, we take these couplings to be small. As a result, the scalar potential related to $\eta$ can be approximated  as:
 \begin{equation}
 V (\eta) \approx  -  \mu^2_\eta \eta^\dagger \eta + \lambda_\eta (\eta^\dagger \eta)^2\,. 
 \end{equation}
The VEV of $\eta$ can be determined as $v_\eta = \sqrt{\mu^2_\eta/\lambda_\eta}$.

Based on the gauge symmetry, the scalar potential related to the scalar fields, such as $H$, $S^\ell$, and $S$,  are written  as:  
 \begin{align}
 V & = -\mu^2_H  H^\dagger H + \lambda_H (H^+ H)^2 + \mu^2_S S^\dagger S + \lambda_S (S^\dagger S)^2  + \sum_{\ell=e,\mu} \left( \mu^2_{S^\ell} S^{\ell \dagger} S^\ell +  \lambda_{S^\ell} (S^{\ell \dagger} S^\ell)^2  \right)  \nonumber \\
 & + \lambda_{HS} H^\dagger H S^\dagger S+ \sum_{\ell=e,\mu} \left( \lambda_{HS^\ell} H^\dagger H + \lambda_{SS^\ell} S^\dagger S \right) S^{\ell\dagger} S^{\ell}  + \lambda_{\mu e} (S^e S^{e\dagger}) (S^\mu S^{\mu\dagger}) \nonumber \\
 & + \left[ \mu^2_{\mu e} S^e S^\mu + \mu_{ee} (S^e)^2 S^\dagger 
+ \mu_{\mu \mu} (S^\mu)^2 S + \tilde{\mu}_{\mu e} S^e S^{\mu \dagger} S^\dagger 
  + \lambda^{\mu e}_H H^\dagger H S^e S^\mu \right. \nonumber \\
  & \left. + \lambda^{e\mu}_S S^e S^\mu S^\dagger S + \sum_{\ell=e,\mu} \lambda^{e\mu}_{S^\ell} S^e S^\mu S^{\ell\dagger} S^{\ell} + \lambda'_{\mu e} (S^e S^\mu)^2 + H.c. \right] \label{eq:Vp}
 \end{align}
 Using the neutral scalar fields, which are expanded around their VEVs and defined as:
  \begin{align}
  H^0 = \frac{v+\phi }{\sqrt{2}}\, , ~ S=\frac{v_S + s}{\sqrt{2}}\,, ~ S^\ell= \frac{v_{S^\ell} + s^\ell }{\sqrt{2}}\,,
  \end{align}
 the minimal conditions of the VEVs can be found as: 
 \begin{align}
 v^2 & = \frac{\mu^2_H}{\lambda_H} - \frac{1}{2\lambda_H} \left( \lambda_{HS} v^2_S + 2 \lambda^{\mu e}_H v_{S^e} v_{S^\mu} +\sum_{\ell} \lambda_{HS^\ell} v^2_{S^\ell} \right)\,,  \nonumber \\
 v^2_S & = -\frac{1}{\lambda_S}\left(\mu^2_S + \frac{\lambda_{HS} v^2 + \lambda_{SS^\ell} v^2_{S^\ell}}{2} + \lambda^{\mu e}_S v^e_S v^\mu_S \right) -\frac{1}{\sqrt{2} \lambda_S  } \left( \sum_\ell \mu_{\ell \ell} v^2_{S^\ell} + \tilde{\mu}_{\mu e} v_{S^e} v_{S^\mu}\right) \,, \nonumber \\
 \lambda_{S^e} v^3_{S^e} & = - \left(\mu^2_{S^e} + \frac{\lambda_{\mu e} + 2\lambda'_{\mu e}}{2} v^2_{S^\mu} + \frac{\lambda_{HS^e} v^2 + \lambda_{HS^e} v^2_S}{2} + \sqrt{2} \mu_{ee} v_S  + \frac{3 \lambda^{\mu e}_{S^e} }{2} v_{S^e} v_{S^\mu} \right) v_{S^e}  \nonumber \\
 & - \frac{1}{2}\left( \lambda^{\mu e}_S v^2_S + \lambda^{\mu e}_H v^2 + 2\mu^2_{\mu e} + \sqrt{2} \tilde{\mu}_{\mu e} v_S  + \lambda^{\mu e}_{S^\mu} v^2_{S^\mu}  \right)v_{S^\mu} \,,  \nonumber \\
  \lambda_{S^\mu} v^3_{S^\mu} & = - \left(\mu^2_{S^\mu} + \frac{\lambda_{\mu e} + 2\lambda'_{\mu e}}{2} v^2_{S^e} + \frac{\lambda_{HS^\mu} v^2 +\lambda_{SS^\mu} v^2_S}{2} + \sqrt{2} v_S \mu_{\mu\mu} + \frac{3 \lambda^{\mu e}_{S^\mu} }{2} v_{S^e} v_{S^\mu}\right) v_{S^\mu}  \nonumber \\
 & - \frac{1}{2}\left( \lambda^{\mu e}_S v^2_S + \lambda^{\mu e}_H v^2 + 2 \mu^2_{\mu e} + \sqrt{2} \tilde{\mu}_{\mu e} v_S  + \lambda^{\mu e}_{S^e} v^2_{S^e}\right)v_{S^e}  \,, \label{eq:VEVs}
 \end{align}
 where we have used $\partial V/\partial v_i =0$ with $v_i= v, v_S$, and  $v_{S^\ell}$. In addition, the symmetric  scalar mass-square matrix is obtained as:
  \begin{equation}
  m^2_S = \left(
\begin{array}{cccc}
  m^2_{S^e}  & m^2_{s^e s^\mu}  & m^2_{s^e \phi}  &  m^2_{s^e s}  \\
 m^2_{s^e s^\mu}  & m^2_{S^\mu}  & m^2_{s^\mu \phi}  &  m^2_{s^\mu s}    \\
 m^2_{s^e \phi}  &  m^2_{s^\mu \phi} &  m^2_{\phi} &  m^2_{\phi s}  \\
 m^2_{s^e s} &  m^2_{s^\mu s}   & m^2_{\phi s}  & m^2_{S}  \\
 \end{array}
\right)\,,
  \end{equation}
  where the matrix elements are obtained as:
  \begin{align}
  m^2_{S^e} & = 2 \lambda_{S^e} v^2_{S^e} + \frac{3}{2} \lambda^{\mu e}_{S^e} v_{S^e} v_{S^\mu} \nonumber \\
  & - \frac{1}{2} \left( \lambda^{\mu e}_S v^2_S + \lambda^{\mu e}_H v^2 + \lambda^{\mu e}_{S^\mu} v^2_{S^\mu}+ 2 \mu^2_{\mu e} +\sqrt{2} \tilde{\mu}_{\mu e} v_S  \right) \frac{v_{S^\mu}}{v_{S^e}} \,, \nonumber \\
  m^2_{S^\mu} & = 2 \lambda_{S^\mu} v^2_{S^\mu} + \frac{3}{2} \lambda^{\mu e}_{S^\mu} v_{S^e} v_{S^\mu} \nonumber \\
  & - \frac{1}{2} \left( \lambda^{\mu e}_S v^2_S + \lambda^{\mu e}_H v^2 + \lambda^{\mu e}_{S^e} v^2_{S^e}+ 2 \mu^2_{\mu e} +\sqrt{2} \tilde{\mu}_{\mu e} v_S  \right) \frac{v_{S^e}}{v_{S^\mu}} \,, \nonumber \\
  m^2_{\phi} & = 2 v^2 \lambda_H\,, ~~ m^2_{S} = 2 \lambda_S v^2_S  - \frac{1}{\sqrt{2} v_S} \left( \sum_{\ell} \mu_{\ell \ell} v^2_{S^\ell} + \tilde{\mu}_{\mu e} v_{S^e} v_{S^\mu}\right)\,, \nonumber 
  \end{align}
  \begin{align}
  %
m^2_{s^e s^\mu} & = (2 \lambda_{S^e} v^2_{S^e} -m^2_{S^e}) \frac{v_{S^e}}{v_{S^\mu}} + (\lambda_{\mu e} + 2\lambda'_{\mu e}) v_{S^e} v_{S^\mu} + \frac{3}{2} \sum_\ell \lambda^{\mu e}_{S^\ell} v^2_{S^\ell} \,, \nonumber \\
  m^2_{s^\ell \phi} & =( \lambda^{\mu e}_{H}  r_v + \lambda_{HS^\ell}  ) vv_{S^\ell}\,, ~~  m^2_{\phi S} = \lambda_{HS} v v_{S} \,, \nonumber \\
  m^2_{s^{e(\mu)} s} &= \sqrt{2} \mu_{e e(\mu \mu)} v_{S^{e(\mu)}}+ \left( \lambda^{\mu e}_{S} v_S  +\frac{\tilde{\mu}_{\mu e} }{\sqrt{2}} \right)v_{S^{\mu (e)}}\,.
  \end{align}
In $m^2_{s^\ell \phi}$, $r^\ell_v=v_{S^\mu}/v_{S^e}(v_{S^e}/v_{S^\mu})$ for $\ell=e(\mu)$.  The relations in Eq.~(\ref{eq:VEVs}) have been applied to $m^2_{S^\ell}$, $m^2_\phi$, and $m^2_S$. 
  
  To explain the anomalous lepton $g-2$, we numerically find  that $m_{S^\ell} < 100$ GeV, $v_{S^\ell}\lesssim 1 ~{\rm GeV}$, and ($\lambda^{\mu e}_{H}, \lambda_{HS^\ell}) \gtrsim 0$ are preferred in the model.  In addition,  to fit the neutrino data, we need $\mu^2_{\mu e} \ll v^2$ when the Yukawa couplings are taken to be of $O(10^{-5}-10^{-4})$. Thus, the positive  $v$, $v_{S, S^\ell}$, and $m^2_{S^\ell,\phi, S}$ can be achieved when the parameters are taken to follow the conditions:
   \begin{align}
    & \lambda_{H, S, S^\ell}>0, ~ \mu^2_{H,S^\ell, S} >0\,, ~  \mu^2_S + \frac{\lambda_{HS} v^2}{2} <  0\,, ~\lambda_{HS} < 0\,, \nonumber \\
    &  \lambda^{\mu e}_{S}  < 0 \,,  ~ 0< 2\mu^2_{\mu e} + \lambda_{HS^\ell} v^2  < -(\lambda^{\mu e}_S v^2_S + \lambda^{\mu e}_{H} v^2 + \sqrt{2} \tilde{\mu}_{\mu e} v_S)\,. \label{eq:para_c}
  \end{align}
   In order to obtain the stable vacuum and the scalar potential bounded from below, we further require that the quartic parameters in the scalar potential should satisfy the conditions~\cite{Kannike:2016fmd}:
   \begin{align}
   & a_1=\lambda_{HS} + \sqrt{\lambda_H \lambda_S} > 0\,, ~ a_2=\lambda_{HS^\ell} + \sqrt{\lambda_H \lambda_{S^\ell} }>0\,, ~ a_3=\lambda_{S S^\ell} + \sqrt{\lambda_{S} \lambda_{S^\ell}}>0\nonumber \\
   & \sqrt{\lambda_H \lambda_S \lambda_{S^\ell}} + \lambda_{HS} \sqrt{\lambda_{S^\ell}} + \lambda_{HS^\ell} \sqrt{\lambda_S}+\lambda_{SS^\ell} \sqrt{\lambda_H} + \sqrt{2 a_1 a_2 a_3} > 0 \,. \label{eq:posi}
   \end{align}
  In addition, to avoid the strict constraint from the precision Higgs measurements, $m^2_{s^\ell \phi} \ll m^2_{\phi}$ is necessary; that is, 
  \begin{equation}
  (\lambda^{\mu e}_H r^\ell_v + \lambda_{HS^\ell} ) v_{S^\ell} \ll 2 v \lambda_H\,.
  \end{equation}
   Using these conditions, it can be  found that with $|\lambda_{HS}| < 1$,  the scalar $\phi$ can approximate the SM Higgs $h$ with a mass of $m_h=125$ GeV.   To numerically illustrate the scalar masses, we take the parameter values, which obey the conditions shown in Eqs.~(\ref{eq:para_c}) and (\ref{eq:posi}), as: $(v,v_S,v_{S^e}, v_{S^\mu})=(246, 100,1,2)$ GeV, $\tilde{\mu}_{\mu e}=-57$ GeV, $(\lambda_H, \lambda_S)=(0.14, 8)$, and $(\lambda_{HS}, \lambda_{HS^{e}},\lambda_{HS^\mu},\lambda^{\mu e}_S, \lambda^{\mu e}_H) =(-0.5, 1.6, 2.0, -0.5, 0.05$). The resulting values for $m^2_S$ are given as:
 \begin{align}
 m^2_{S^e} & \approx m^2_{S^\mu}  \approx -\left( \lambda^{\mu e}_S v^2_S + \lambda^{\mu e}_H v^2 + \sqrt{2} \tilde{\mu}_{\mu e} v_S \right)\sim 90^2\,, \nonumber \\
 m^2_\phi & = 2\lambda_H v^2 \sim 129^2\,, ~ m^2_{S}  \approx 2 \lambda_S v^2_S \sim 400^2\,,  ~ m^2_{s^e s^\mu}\approx -m^2_{S^e}/2 \sim - 90^2/2 \,, \nonumber \\
 m^2_{s^{e(\mu)} \phi} & \approx \lambda_{H S^{e(\mu)}} v v_{S^{e(\mu)}}\sim 394( 1000)\,, ~ m^2_{\phi S} = \lambda_{HS} v v_S \sim  - 1.23\times 10^4\,, \nonumber \\
  m^2_{s^{e(\mu) s}} & \approx \left( \lambda^{\mu e}_S v_S + \tilde{\mu}_{\mu e}/\sqrt{2} \right) v_{S^{\mu(e)}} \sim -180(-90)\,. 
 \end{align}
Accordingly, the mass eigenvalues in units of GeV can be obtained as: $m_{\bar s}\approx 401.3$, $m_h\approx 125.4$, $m_{\bar s^\mu}\approx 109.6$, and $m_{\bar s^e}\approx 63.6$, where the corresponding eigenstates of $s$ and h can be expressed as:
 \begin{equation}
 \bar s \approx 0.996 s -0.085 \phi\,, ~ h \approx -0.084 s - 0.986 m_\phi - 0.141 s^\mu \,. \label{eq:s_mixing}
 \end{equation}
 It can be seen that $h$ still aligns $m_\phi$, and we can suppress the pair production for $h\to s^\ell s^{\ell'}$ by taking proper parameter values, .

  After $U(1)_{e-\mu}$ symmetry is spontaneously broken,  the associated  $Z'$-gauge boson becomes a massive particle and its mass  can be  obtained as:
  \begin{equation}
  m_{Z'} = g_{Z'} \sqrt{v^2_\eta+4 v_S^2 + v_{S^e}^2 + v_{S^\mu}^2},
  \end{equation}
  where $g_{Z'}$ is $U(1)_{e - \mu}$ gauge coupling constant.

 \section{ Charged lepton and neutrino mass matrices}
 
\begin{figure}[phtb]
\begin{center}
\includegraphics[scale=0.65]{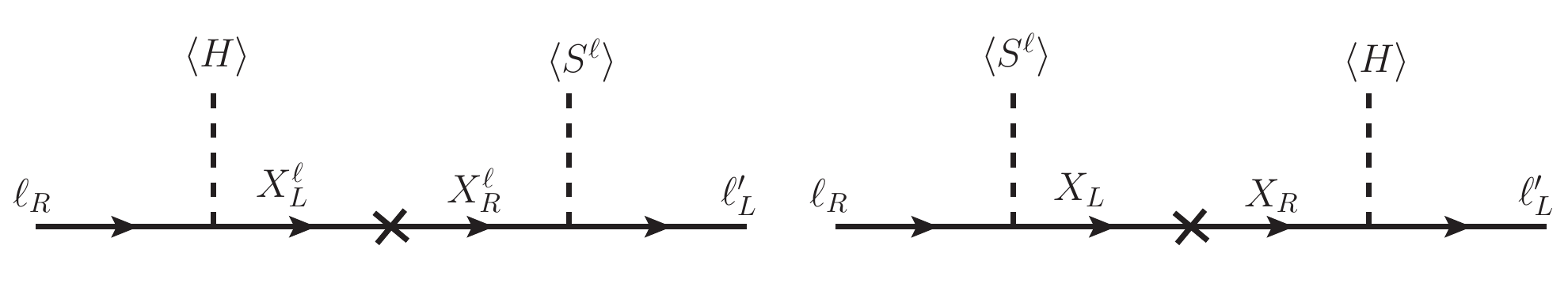}
 \caption{ Feynman diagrams used to induce the electron and muon masses.}
\label{fig:mass_emu}
\end{center}
\end{figure}

  From Eq.~(\ref{eq:Yukawa}), it can be seen that with the exception of the $\tau$-lepton, the electron and muon do not directly obtain their masses with the Higgs mechanism.  Nevertheless, their masses can be induced through the mixings with $X^\ell_{L,R}$ and $X_{L,R}$, where the Feynman diagrams are shown in Fig.~\ref{fig:mass_emu}.
  
 Using the Yukawa couplings and the VEVs of scalar fields, the $6\times 6$ charged lepton mass matrix in the flavor basis of $(e, \mu, \tau, X^e, X^\mu, X)_{L,R}$ is written as:
 \begin{equation}
M_{CL}=\left(
\begin{array}{cccccc}
  0 & 0  & 0  &  m_{eX^e} & m_{e X^\mu} & m_{e X} \\
  0 & 0  & 0  &  m_{\mu X^e} & m_{\mu X^\mu}  & m_{\mu X}  \\
 0 &  0 &  m^0_{\tau} &  m_{\tau X^e} & m_{\tau X^\mu}  & m_{\tau X} \\
 m_{X^e e} & 0   & 0  & M_{X^e} &  0  & 0 \\
  0  & m_{X^\mu \mu}   & 0  & 0  & M_{X^\mu}  &  0 \\
 m_{X e} &  m_{X\mu}  &  0 & 0 &  0 & M_{X}  \\
\end{array}
\right)\,, \label{eq:MCL}
 \end{equation}
where the matrix elements are given as:
 \begin{align}
 m_{\ell' X^\ell} & = \frac{y^{\ell}_{\ell'} v_{S^\ell} }{\sqrt{2}} \,, ~ m_{\ell' X}  = \frac{\tilde{y}_{\ell'} v }{\sqrt{2} }\,, \nonumber \\
 m_{X^{\ell} \ell} & = \frac{y^\ell_X v}{\sqrt{2}}\,, ~ m_{X \ell} = \frac{y_\ell v_{S^\ell}}{\sqrt{2}}\,. \label{eq:mlXl}
 \end{align}
Since several phenomena are related to $m_{\ell' X^\ell(X)}$ and  $m_{X^\ell(X) \ell}$, we thus use them as the free parameters instead of the corresponding Yukawa couplings and VEVs.  In this study, we assume that the relevant Yukawa couplings are real parameters.  The mass matrix $M_{CL}$ can be diagonalized by the  unitary matrices $U_{L, R}$ as $M^{\rm diag}_{CL} = U_{L} M_{CL} U^\dagger_R$. The $m_{\ell}$ eigenvalues can be obtained using  $M^{\rm diag}_{CL} M^{\rm diag\dagger}_{CL}= U_L M_{CL} M^\dagger_{CL} U^\dagger_L$ and $M^{\rm diag \dagger}_{CL} M^{\rm diag}_{CL}= U_R M^{\dagger}_{CL} M_{CL} U^\dagger_R$. 

\begin{figure}[phtb]
\begin{center}
\includegraphics[scale=0.65]{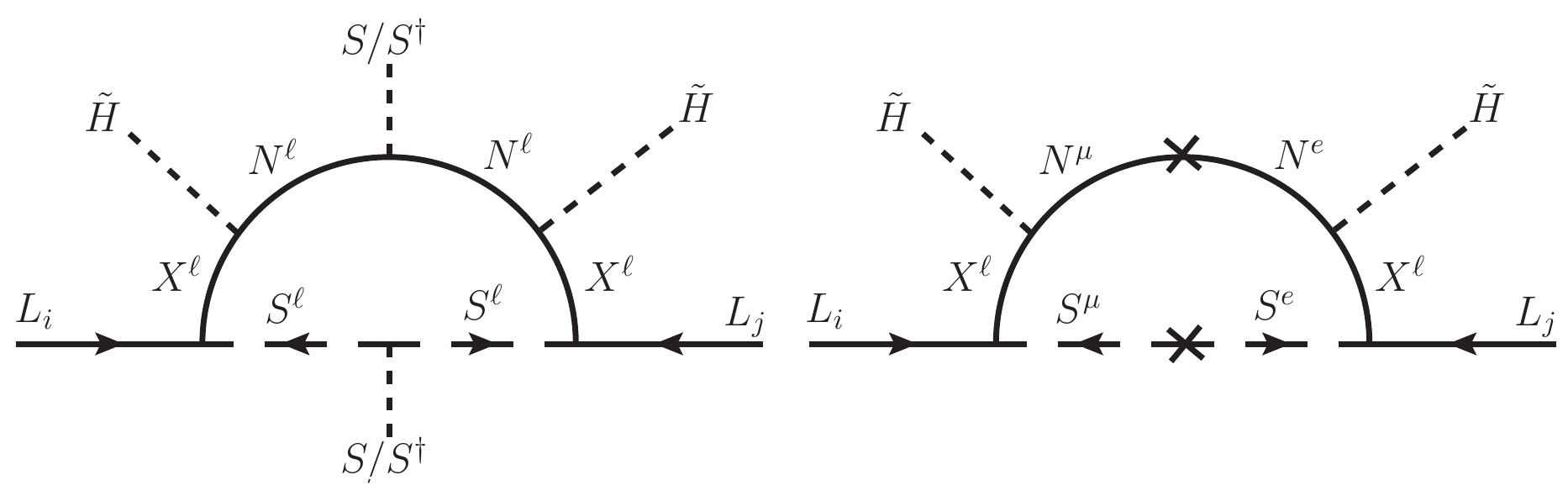}
 \caption{ Feynman diagrams used to generate the neutrino masses.}
\label{fig:neutrino}
\end{center}
\end{figure}

Although the right-handed neutrinos $N^\ell$ are introduced, since the left-handed SM leptons do not carry the $U(1)_{e-\mu}$ charges, the neutrino mass cannot be generated at the tree-level in the model. Nonetheless, the neutrino mass can be produced through the radiative effects, for which the Feynman diagrams are shown in Fig.~\ref{fig:neutrino}.

 Using the Yukawa and scalar couplings shown in Eq.~(\ref{eq:Yukawa}) and Eq.~(\ref{eq:Vp}),  respectively, 
 the loop-induced neutrino mass matrix elements, denoted as $\nu^T_j C m^\nu_{ji} \nu_i$, can be obtained as:
 \begin{align}
 m^\nu_{ji}  &= \frac{v v^2_S }{\sqrt{2} (4\pi)^2} \sum_{\ell=e,\mu} \frac{ \mu_{\ell \ell} m_{N^\ell} ( \tilde{y}^\ell_X)^2}{m^4_{X^\ell} } \frac{m_{j X^\ell} m_{i X^\ell}}{ (v_{S^\ell})^2} J_0 \left( \frac{m^2_{S^\ell}}{m^2_{X^\ell}}, \frac{m^2_{N^\ell} }{m^2_{X^\ell} }\right) \nonumber \\
&  -\frac{\tilde{y}^e_X \tilde{y}^\mu_X v^2 \mu^2_{\mu e} m_{N_{\mu e}} }{(4\pi)^2 v_{S^e} v_{S^\mu}}  \frac{(m_{j X^e} m_{i X^\mu} + m_{jX^\mu}  m_{i X^e}) }{m^4_{X^\ell }} J_{1}\left( \frac{m^2_{S^\ell}}{m^2_{X^\ell}}, \frac{m^2_{N^\mu} }{m^2_{X^\ell} }, \frac{m^2_{N^e} }{m^2_{X^\ell} }\right) \,, \label{eq:mnuji}
 \end{align}
where the Latin letters $j,i$ denote the active neutrino flavors, and the first (second) term originates from the left (right) panel in Fig.~\ref{fig:neutrino}. For simplicity, we take $m_{S^e}=m_{S^\mu}=m_{S^\ell}$ and $m_{X^e}=m_{X^\mu}=m_{X^\ell}$; $m_{N^\ell} = h_\ell v_S/\sqrt{2}$ and $m_{\ell' X^\ell}(m_{\ell' X})$ defined in Eq.~(\ref{eq:mlXl}) are used, and the loop integrals are expressed as:
 \begin{align}
 J_0(a, b) & =  \int^1_0 dx \int^x_0 dy \frac{(1-x)(x-y)}{(1+(a-1) x + (b-a) y)^2}\,, \nonumber \\
 J_1(a, b, c) & = \int^1_0 dx \int^x_0 dy \int^y_0 dz \frac{(1-x) (x-y)}{(1+(a-1) x + (b-a) y + (c-a) z)^3}\,.
 \end{align} 

The neutrino mass matrix can be diagonalized by the Pontecorvo-Maki-Nakagawa-Sakata (PMNS) matrix as:
 \begin{equation}
 m^{\nu}_{ij} = U^*_{MNS} m_\nu^{\rm diag} U^\dagger_{MNS}, \label{eq:mnuij}
 \end{equation}
 where $m_\nu^{\rm diag} = {\rm diag}(m_1, m_2, m_3)$, and 
 the PMNS matrix can be parametrized as~\cite{PDG}:
 \begin{align}
\label{MNS}
U_{\rm MNS} & = \begin{pmatrix} 
c_{12} c_{13} & s_{12} c_{13} & s_{13} e^{-i \delta} \\
-s_{12} c_{23} - c_{12} s_{23} s_{13} e^{i \delta} & c_{12} c_{23} -s_{12} s_{23} s_{13} e^{i \delta} & s_{23} c_{13} \\
s_{12} s_{23} - c_{12} c_{23} s_{13} e^{i \delta} & -c_{12} s_{23} - s_{12} c_{23} s_{13} e^{i \delta} & c_{23} c_{13} 
\end{pmatrix} \nonumber \\
& \times \text{diag}(1, e^{i \alpha_{21}/2}, e^{i\alpha_{31}/2})\,,
\end{align}
with $s_{ij} \equiv \sin \theta_{ij}$ and  $c_{ij} \equiv \cos \theta_{ij}$.  $\delta$ is the Dirac CP violating phase, and $\alpha_{21, 31}$ are  Majorana CP violating phases. 

\section{ $\ell' \to \ell \gamma$, and lepton $g-2$}

\begin{figure}[phtb]
\begin{center}
\includegraphics[scale=0.65]{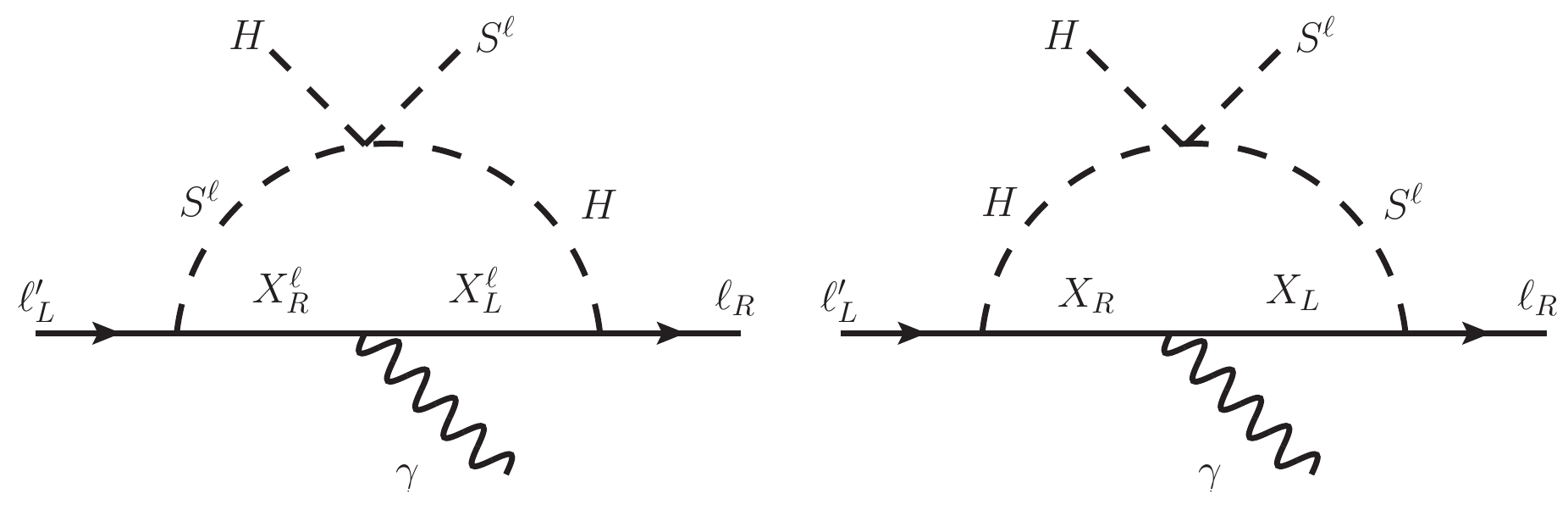}
 \caption{Selected Feynman diagrams for the $\ell'\to \ell \gamma $ decays.}
\label{fig:lep_dipole}
\end{center}
\end{figure}

If we add the couplings $H^\dag H S^{\ell\dag} S^\ell$ and $H^\dag H S^e S^\mu$ to Fig.~\ref{fig:mass_emu}, it can be seen that the radiative LFV processes can be induced through the loop effects, for which the relevant Feynman diagrams are sketched in Fig.~\ref{fig:lep_dipole}.  The current  experimental upper limits on the BR for the relevant LFV processes are given as~\cite{PDG}:
\begin{align} 
BR(\mu \to e \gamma) & < 4.2 \times 10^{-13}\,, \nonumber \\
BR(\tau \to e \gamma)& < 3.3 \times 10^{-8}\,, \nonumber \\
BR(\tau\to \mu \gamma)& < 4.4 \times 10^{-8}\,.
  \end{align}
Since the radiative LFV processes are dominant in the model, we skip the analysis for the subleading $\mu\to 3e$ and $\tau\to 3 \ell$ decays.

Using the Yukawa and scalar couplings, the effective interactions for $\ell'\to \ell\gamma$ can be written  as:
\begin{align}
{\cal L}_{\ell'\to \ell \gamma} = \frac{e }{2} m_{\ell'} \bar \ell \sigma_{\mu \nu} \left( T^{\ell' \ell}_{L} P_L + T^{\ell' \ell}_R P_R \right) \ell' F^{\mu \nu}\,,
\end{align}
where the Wilson coefficients in the model are obtained as:
\begin{align}
T^{\mu e}_L &  = \frac{ \lambda_{HS^e} + \lambda^{\mu e}_{H} r^e_v }{2 m_\mu (4\pi)^2} \left[ \frac{m_{\mu X^e} m_{X^e e} }{m^3_{X^\ell}} J_2\left(\frac{m^2_{S^\ell}}{m^2_{X^\ell}}, \frac{m^2_\phi}{m^2_{X^\ell}} \right)  + \frac{m_{\mu X} m_{X e}}{m^3_X} J_2\left(\frac{m^2_{S^\ell}}{m^2_{X}}, \frac{m^2_\phi}{m^2_{X}} \right)\right] \,,\nonumber \\
T^{\mu e}_R & =  \frac{ \lambda_{HS^\mu} + \lambda^{\mu e}_H r^\mu_v }{2 m_\mu (4\pi)^2} \left[ \frac{m_{e X^\mu} m_{X^\mu \mu}}{m^3_{X^\ell}} J_2\left(\frac{m^2_{S^\ell}}{m^2_{X^\ell}}, \frac{m^2_\phi}{m^2_{X^\ell}} \right) + \frac{m_{e X} m_{X \mu}}{m^3_X} J_2\left(\frac{m^2_{S^\ell}}{m^2_{X}}, \frac{m^2_\phi}{m^2_{X}} \right)\right] \,, \nonumber \\
T^{\tau \ell} _L & = \frac{\lambda_{HS^\ell}+\lambda^{\mu e}_{H} r^\ell_v}{2m_\tau (4\pi)^2} \left[ \frac{m_{\tau X^\ell} m_{X^\ell \ell}  }{m^3_{X^\ell}}
J_2\left(\frac{m^2_{S^\ell}}{m^2_{X^\ell}}, \frac{m^2_\phi}{m^2_{X^\ell}} \right) +  \frac{m_{\tau X} m_{X \ell} }{m^3_X} J_2\left(\frac{m^2_{S^\ell}}{m^2_{X}}, \frac{m^2_\phi}{m^2_{X}} \right)\right]\,, \label{eq:FFs}
\end{align}
$T^{\tau \ell}_R=0$, and  the loop integral $J_2$ is defined by:
\begin{equation}
J_2(a,b) = \int^1_0 dx_1 \int^{x_1}_0 dx_2 \int^{x_2}_0 dx_3 \frac{1-x_2}{(1+ (a-1)x_2 + (b-a)x_3 )^2}\,.
\end{equation}
 The definitions shown in Eq.~(\ref{eq:mlXl}) have been  used.  Due to $m_\ell \ll m_{\ell'}$, we have neglected the $m_\ell$ effects.  As a result, the BR for the $\ell'\to \ell \gamma$ decay can be written as: 
\begin{equation}
BR(\ell' \to \ell \gamma) = \tau_{\ell'} \frac{\alpha m^5_{\ell'}}{4} \left( |T^{\ell' \ell}_L |^2 + |T^{\ell' \ell}_R |^2\right)\,.
\end{equation}
 In order to satisfy the current upper limit  of $BR^{\rm exp}(\mu \to e \gamma)< 4.2 \times 10^{-13}$, one can take $m_{\mu X^e}\approx m_{e X^\mu} \approx 0$ or the conditions assumed as:
  \begin{align}
  \frac{m_{\mu X^e} m_{X^e e} }{m^3_{X^\ell}} J_2\left(\frac{m^2_{S^\ell}}{m^2_{X^\ell}}, \frac{m^2_\phi}{m^2_{X^\ell}} \right) \approx - \frac{m_{\mu X} m_{Xe}}{m^3_X} J_2\left(\frac{m^2_{S^\ell}}{m^2_{X}}, \frac{m^2_\phi}{m^2_{X}} \right)\,, \nonumber \\
   \frac{m_{e X^\mu} m_{X^\mu \mu} }{m^3_{X^\ell}} J_2\left(\frac{m^2_{S^\ell}}{m^2_{X^\ell}}, \frac{m^2_\phi}{m^2_{X^\ell}} \right) \approx - \frac{m_{e X} m_{X\mu}}{m^3_X} J_2\left(\frac{m^2_{S^\ell}}{m^2_{X}}, \frac{m^2_\phi}{m^2_{X}} \right)\,. \label{eq:mutoe}
  \end{align}
In this study, we adopt the latter requirements shown in Eq.~(\ref{eq:mutoe}). Hence,  in the numerical analysis, we only focus on the $\tau \to \ell \gamma$ decays. 

It is known that the radiative quantum corrections to a lepton current can be expressed as:
 \begin{equation}
 \Gamma^\alpha = \bar \ell (p') \left[ \gamma^\alpha  F_1(k^2) + \frac{i \sigma^{\alpha \beta} k_\beta}{2m_\ell} F_2(k^2) \right] \ell(p) \,,
 \end{equation}
 where the lepton $g-2$ can  be defined by:
  \begin{equation}
  a_\ell = \frac{g_\ell-2}{2} = F_2(0)\,.
  \end{equation}
 Using this definition, the lepton $g-2$ can be induced by Fig.~\ref{fig:lep_dipole} with $\ell'=\ell$. Based on the results shown in Eq.~(\ref{eq:FFs}), the lepton $g-2$ can be formulated as:
\begin{align}
\delta a_\ell & = m_\ell \frac{\lambda_{HS^\ell} + \lambda^{\mu e}_{H} r^\ell_v }{(4\pi)^2} \left[ \frac{m_{\ell X^\ell} m_{X^\ell \ell} }{m^3_{X^{\ell}}} J_{2}\left( \frac{m^2_{S^\ell}}{m^2_{X^\ell}}, \frac{m^2_\phi}{m^2_{X^\ell} }\right) + \frac{m_{\ell X} m_{X \ell} }{m^3_{X}} J_{2}\left( \frac{m^2_{S^\ell}}{m^2_{X}}, \frac{m^2_\phi}{m^2_{X} }\right) \right]\,. \label{eq:lep_g-2}
%
\end{align}
It can be seen that the obtained $\delta a_\mu$ is proportional to $m_\ell$ and linearly depends on $\lambda_{HS^\ell} + \lambda^{\mu e}_{H} r^\ell_v$, which is related to $m^2_{s^\ell \phi}$.  Since $\lambda_{HS^\ell}$ and $\lambda^{\mu e}_{H}$ are free parameters, to use fewer scanned parameters,  for simplicity, we take $\lambda^{\mu e}_{H} =0$  in our numerical analysis.

 As mentioned in the introduction, the $Z'$-gauge boson can contribute to the lepton $g-2$, and the result can be formulated as: 
 \begin{equation}
\delta  a_{\ell}^{Z'} = \frac{g^2_{Z'} r_\ell^2}{16 \pi^2}  \int_0^1 dx \frac{ x(1-x)(2x-4) -2 r_\ell^2 x^3  }{(1-x)(1-r_\ell^2 x) +r_\ell^2 x}\,,
\end{equation} 
with $r_\ell = m_\ell/m_{Z'}$. Due to the fact that  $2x-4<0$ in the integral, the resulting $\delta a^{Z'}_\ell$ is always negative.  Because $\delta a^{Z'}_{e}/\delta a^{Z'}_{\mu} \approx m^2_e/m^2_\mu \approx 2.4 \times 10^{-5}$, if $\delta a^{Z'}_e \sim - 5\times 10^{-13}$ is taken,  we obtain $\delta a^{Z'}_\mu \sim -2.1 \times 10^{-8}$. However, the large negative $\delta a^{Z'}_{\mu}$ contradicts to the current data, and the sign cannot be  flipped  via other effects in the model. To avoid this issue, we can take proper values for $g_{Z'}$ and $m_{Z'}$ to suppress $\delta a^{Z'}_\mu$. For instance, with $g_{Z'}\sim 5\times 10^{-4}$ and $m_{Z'}\sim 1$ GeV, we have $\delta a^{Z'}_{\mu} \sim - 2.3\times 10^{-11}$; thus, the result will not affect the contributions from Eq.~(\ref{eq:lep_g-2}).



\section{Numerical analysis}

\subsection{ Constraints and setting the scanned parameter regions}

Since $m_e$ and $m_\mu$ are induced through the diagonalization of the $6\times 6$ $M_{CL}$ matrix, basically,  the parameters in $M_{CL}$ have to obtain $m_e\sim 5.1\times 10^{-4}$ GeV and $m_\mu \sim 0.105$ GeV. However, the parameter scan is inefficient when we fit the mass hierarchy  between the electron and the muon. In order to obtain more allowed sampling points, we take  $m_e = (8, 4)\times 10^{-4}$ GeV and $m_\mu=(0.107, 0.103)$ GeV as the constraints.

Although the neutrino mass order  is not yet determined, since  other analyses are not sensitive to the mass order, we use the normal ordering (NO) scheme, i.e. $m_1< m_2 \ll  m_3$, in our study. 
Based on the neutrino oscillation data~\cite{PDG},  the central values of $\theta_{ij}$, $\delta$, and $\Delta m^2_{ij}=m^2_i - m^2_j$ using the global fit can then be obtained as~\cite{deSalas:2017kay}:
\begin{align}
&  \theta_{12} = 34.5^{\circ}\,, ~\theta_{23}=47.7^{\circ}\,, ~\theta_{13} =8.45^{\circ}\,, ~ \delta=218^{\circ}\,,\nonumber \\
&  \Delta m^2_{21} =7.55 \times 10^{-5}\, {\rm eV^2}\,, ~ \Delta m^2_{31} = 2.50\times 10^{-3}\, {\rm eV^2}\,, 
\end{align}
  where $m_{1}=0$ is used, and the Majorana phases are taken to be $\alpha_{21(31)}=0$. Using the $3\sigma$ uncertainties that are shown in \cite{deSalas:2017kay} and the relation shown in Eq.~(\ref{eq:mnuij}),  the  $|m^\nu_{ji}|$ ranges in units of eV  can be estimated as:
\begin{align}
\begin{pmatrix} |m^\nu_{ee}| & |m^\nu_{e\mu}| & |m^\nu_{e\tau}| \\ |m^\nu_{\mu e}| & |m^\nu_{\mu\mu}| & |m^\nu_{\mu\tau}| \\ |m^\nu_{\tau e}| & |m^\nu_{\tau\mu}| & |m^\nu_{\tau\tau}| \end{pmatrix}_{\rm NO} & 
\simeq \begin{pmatrix} 0.11-0.45 & 0.12-0.82 & 0.12-0.82 \\  0.12-0.82 & 2.4-3.3 & 2.0-2.2 \\ 0.12-0.82 & 2.0-2.2  & 2.2-3.1\end{pmatrix} \times 10^{-2}~{\rm eV}\,. \label{eq:v_nu_mass}
\end{align}
We thus use the results in Eq.~(\ref{eq:v_nu_mass}) as the inputs to constrain the free parameters. 

In order to scan the free parameters and obtain the allowed parameter regions when the considered constraints are satisfied, we choose the  parameters in units of GeV from the Yukawa sector as:
 \begin{align}
  m_{e X^e(e X)} & =(3, 12)\,, ~ m_{e X^\mu, \mu X^e} =(-2, 2)\,, ~ m_{\mu X^\mu} =(1, 5)\,,   \nonumber \\
m_{X^e e} & =(-10,-3)\,, ~m_{X^\mu \mu}  =(20, 50)\,, ~ m_{Xe} =(3, 10)\,, \nonumber \\
 ~ m_{N_{\mu e}}& =(-100, 100)\,, ~ m_{\tau X}  =(-10, 10)\,, ~  m_{X^\ell}  =(600, 1200)\,,   \nonumber \\
  m_X & =(800, 1200)\,, ~m_{N^\mu}  = (10, 30)\,, ~ m_{N^e}=(100, 300)\,, 
 \end{align}
and $m_{\tau X^e, \tau X^\mu}  =3$ GeV, whereas  $m_{\mu X}$ and $m_{X\mu}$ are determined by Eq.~(\ref{eq:mutoe}). We note that in order to obtain $\delta a_e <0$ and $\delta a_\mu >0$, we fix  $m_{X^e e}<0$ and $m_{X^\mu \mu}>0$.  The  mass of a vector-like lepton doublet in the range of $120-790$ GeV is excluded by the CMS experiment~\cite{Sirunyan:2019ofn}  in the multilepton final states at $\sqrt{s}=13$ TeV.  Since the $X^\ell-\tau$ mixings are small in our model, the $m_{X^\ell}$ constraint through the coupling $X^\ell \tau Z$ can be looser. The current upper limit on the vector-like lepton singlet is $m_{X} \lesssim 176$ GeV, which was reported  by  ATLAS~\cite{Aad:2015dha}. Hence, the chosen regions for $m_{X^\ell, X}$ follow the current LHC results.  For numerically illustrating the charged lepton masses, we take specific values for the parameters in Eq.~(\ref{eq:MCL}) as:
 \begin{equation}
M_{CL}=\left(
\begin{array}{cccccc}
  0 & 0  & 0  &  8.04 & 0.35. & 8.60\\
  0 & 0  & 0  &  0.45. & 2.52  & 1.55 \\
 0 &  0 &  1.776 &  2 & 2 & 4.29 \\
-3.42 & 0   & 0  & 630 &  0  & 0 \\
  0  & 23.67   & 0  & 0  &630  &  0 \\
5.89  &  -4.59  &  0 & 0 &  0 & 1200 \\
\end{array}
\right)\,.
 \end{equation}
The diagonalized mass eigenvalues  in units of GeV from light to heavy mass can be obtained as:  $m_e\approx 5.5\times 10^{-4}$, $m_\mu\approx 0.1056$, $m_\tau\approx 1.77$7,  $m_{\bar X^\ell}\approx 630$, and $m_{\bar X}\approx 1200$. The associated mass eigenstates for $X^\ell_R$ and $X_L$ are given as:
\begin{align}
\bar X^e_R& \approx 0.005\, e_R + 0.011\, X^\mu_R+ 0.9996 \, X^e_R\,, ~ \bar X^\mu_R \approx  0.04\, \mu_R +0.011 \, X^e_R+ 0.9994\, X^\mu_R\,, \nonumber \\
\bar X_L& \approx 0.007 \, e_L + 0.0013\,  \mu_L +0.0035\, \tau_L + 0.99997 X_L\,. \label{eq:X_mixing}
\end{align}
It can be seen that the $m_e$ and $m_\mu$ results can match the data, and the flavor mixings between the heavy new lepton and the SM lepton are small. 

The massive parameter regions  from the scalar potential are taken as: $v=246$ GeV, $v_S=100$ GeV, $v_{S^{e(\mu)}}=1(2)$ GeV, $m_{S^\ell}=100$ GeV, and:
 \begin{equation}
 \mu_{ee} =(-5,5)~{\rm GeV}\,, ~ \mu_{\mu \mu}=(-1,1)~{\rm GeV}\,, ~ \mu_{\mu e} = (-10, 10) ~{\rm GeV} \,. 
 \end{equation}
The involving dimensionless Yukawa and scalar couplings are set as: $\lambda_{HS^e}=6$, $\lambda_{HS^\mu}=10$,  $\tilde{y}_X^e  = (-2 ,2) \times 10^{-5}$, and  $\tilde{y}_X^\mu  = (-2,2)\times 10^{-4}$. In addition, in order to obtain the sizable $\delta a_{e, \mu}$, we require:
 \begin{align}
 \delta a_e & =( -12.4, -5.2)\times 10^{-13}\,, \nonumber \\
 \delta a_{\mu} & > 5 \times 10^{-10}\,. 
 \end{align}

\subsection{Numerical analysis and discussion}

 From Eq.~(\ref{eq:v_nu_mass}), it can be seen that the matrix elements of $m^\nu_{ji}$ are similar in terms of order of magnitude; thus, we use $2\times 10^8$ sampling points to scan the relevant parameters. However, to obtain the hierarchical values for $m_e$ and $m_\mu$ from the matrix in Eq.~(\ref{eq:MCL}), we use $10^9$ sampling points. 

\begin{figure}[phtb]
\begin{center}
\includegraphics[scale=0.45]{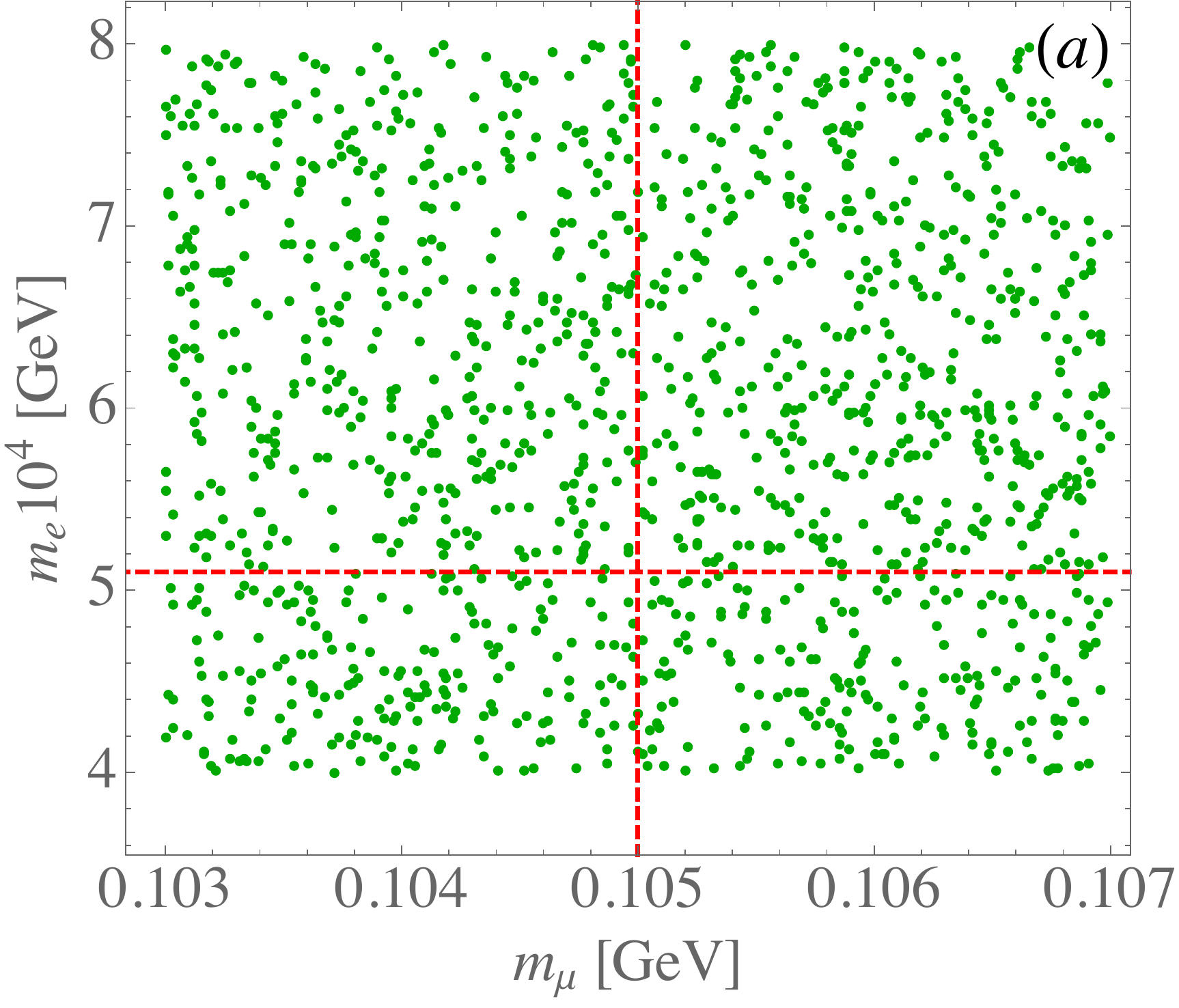}
\includegraphics[scale=0.45]{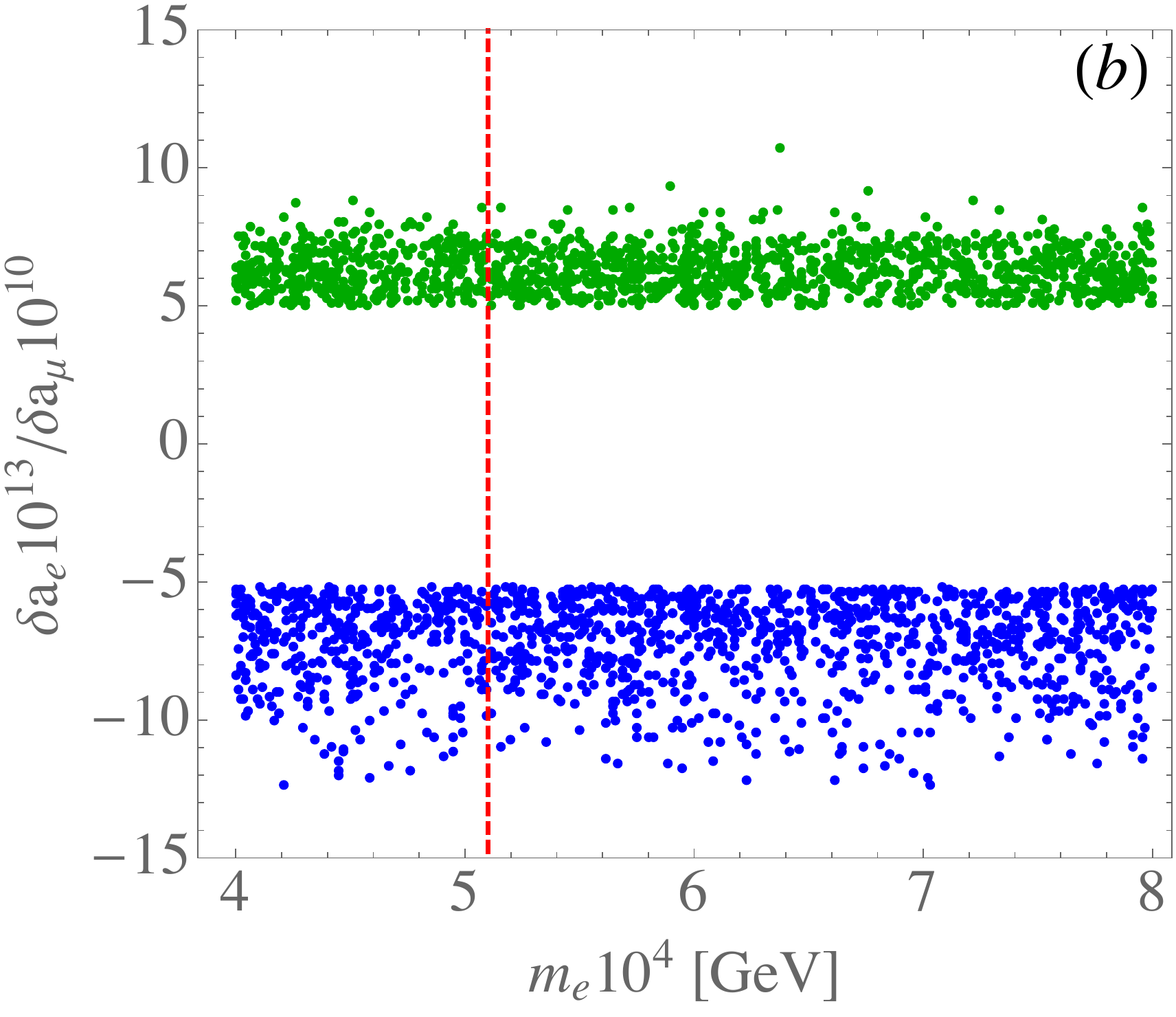}
 \caption{(a) Correlation between the obtained $m_e$ and $m_\mu$, and (b) correlation between $\delta a_{e(\mu)}$ and $m_e$, where $m_e$ is scaled by $10^{-4}$, and $\delta a_{e(\mu)}$ is in units of $10^{-13}(10^{-10})$, indicated by blue(green) points. }
\label{fig:massemua}
\end{center}
\end{figure}

\begin{figure}[phtb]
\begin{center}
\includegraphics[scale=0.5]{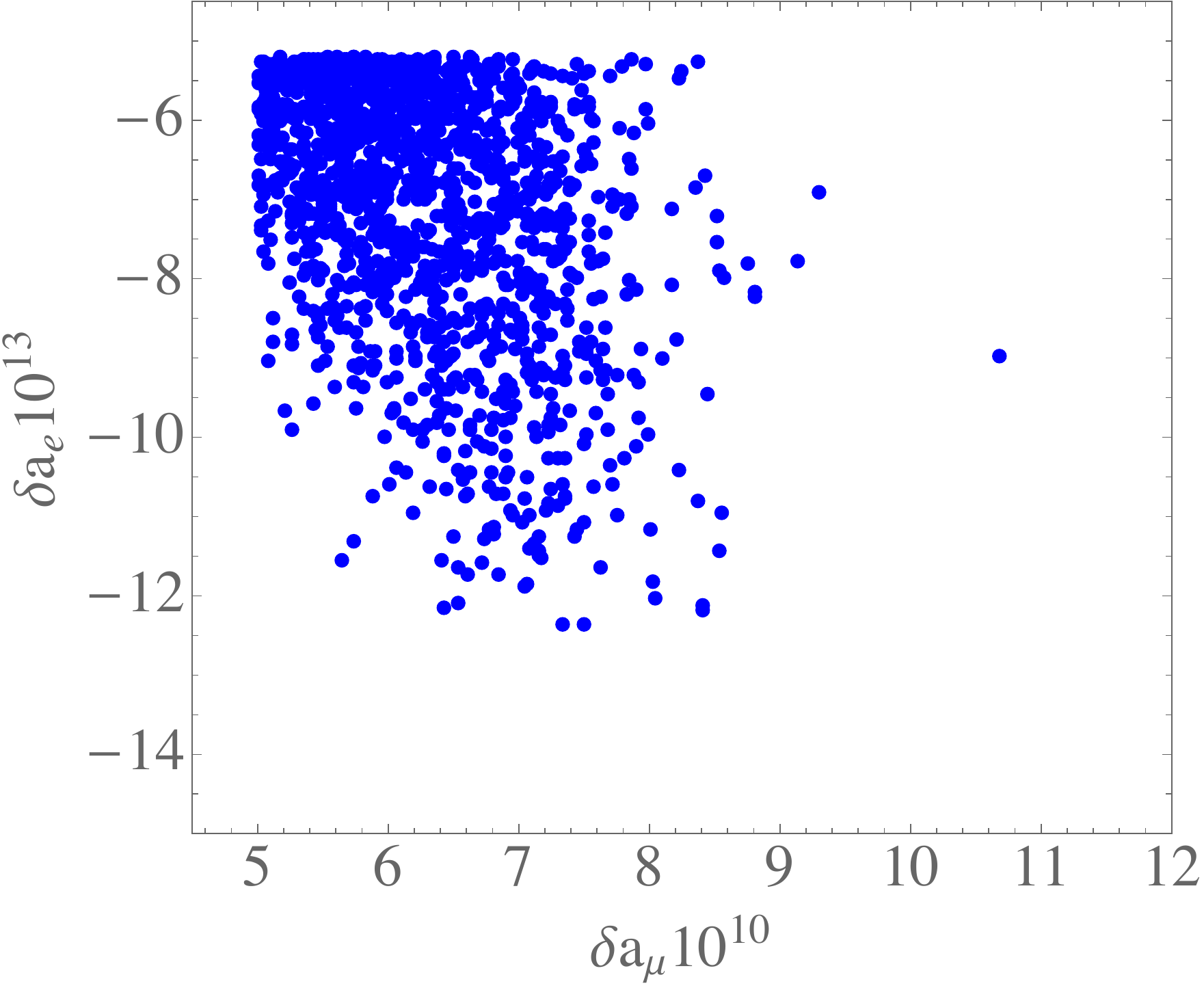}
 \caption{ Correlation between the obtained $\delta a_e$ and $\delta a_\mu$.  }
\label{fig:ae_amu}
\end{center}
\end{figure}

 To show that  $m_e\sim 5.1\times 10^{-4}$ GeV and $m_\mu\sim 0.105$ GeV can be achieved in the chosen parameter regions,  the correlation between the obtained $m_e$ and $m_\mu$ is shown in Fig.~\ref{fig:massemua}(a).  The correlation between  $\delta a_{e(\mu)}$ and $m_e$ can be found in Fig.~\ref{fig:massemua}(b), where  $\delta a_{e(\mu)}$ is in units of $10^{-13}(10^{-10})$ and $m_e$ is scaled by $10^{-4}$,  indicated by blue(green) points. It can be seen that when $m_e\sim 5.1\times 10^{-4}$ GeV is obtained, and the same parameter values can lead to $\delta a_e \sim  O(-10^{-12})$ and $\delta a_\mu\sim O(10^{-9})$.  For clarity, we also show the correlation between the obtained $\delta a_e$ and $\delta a_\mu$ in Fig.~\ref{fig:ae_amu}.

\begin{figure}[phtb]
\begin{center}
\includegraphics[scale=0.5]{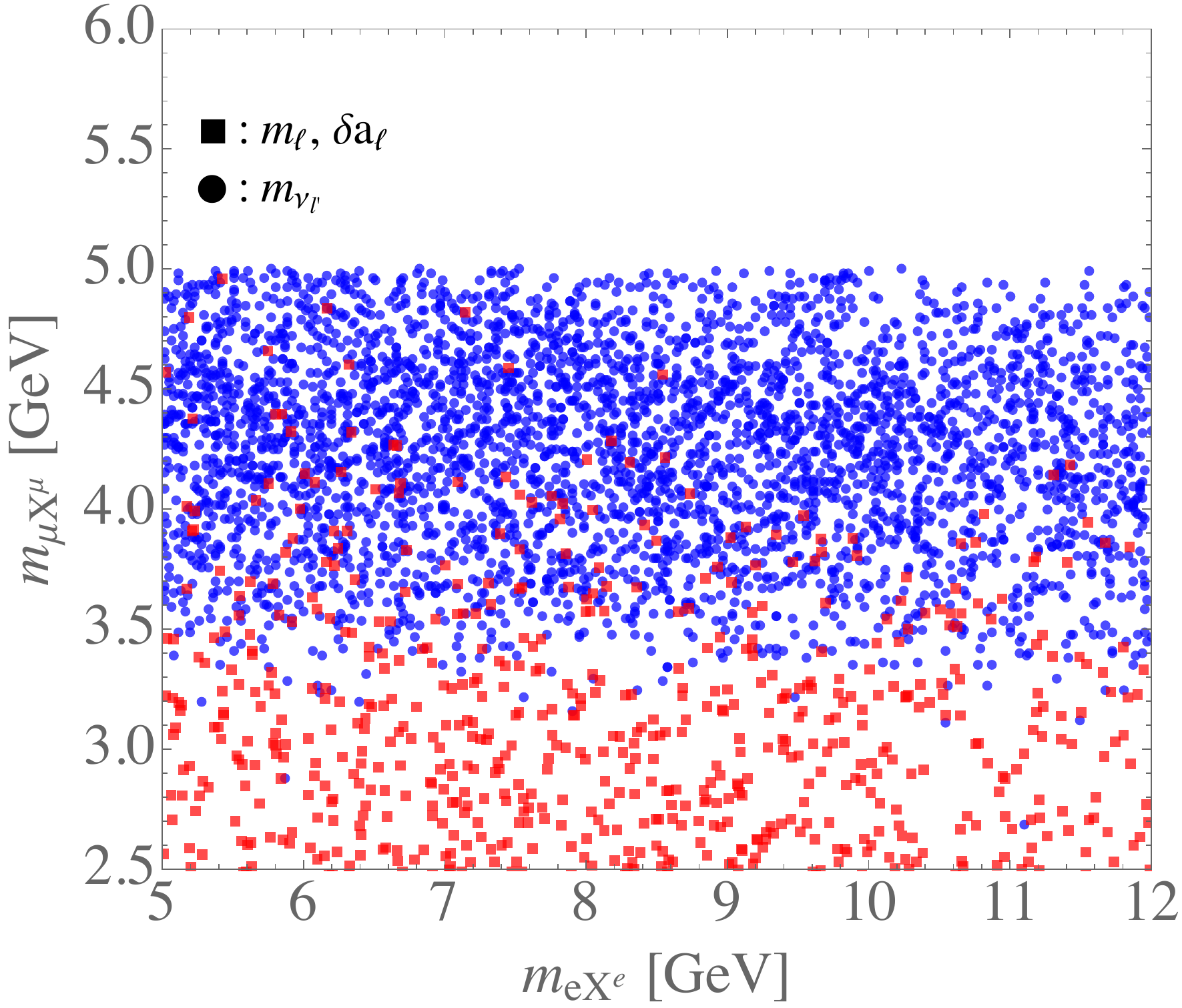}
 \caption{ Resulting correlations for the scanned parameters from different phenomena when they satisfy the chosen ranges, where the filled circles arise from the neutrino data, and the squares are derived from the selected  $m_\ell$ and $\delta a_\ell$ ranges.  }
\label{fig:meXe}
\end{center}
\end{figure}

It is known that $M_{CL}$ in Eq.~(\ref{eq:MCL}), $\delta a_\ell$ in Eq.~(\ref{eq:lep_g-2}), and $m^\nu_{ji}$ in Eq.~(\ref{eq:mnuji}) have common free parameters, such as $m_{\ell X^\ell}$ and $m_{\tau X^\ell}$. To  more efficiently obtain  the allowed parameter regions, we separately scan the parameters to fit the chosen ranges of $m_\ell$ and $\delta a_\ell$ and  the $m^\nu_{ji}$ shown in Eq.~(\ref{eq:v_nu_mass}).  We demonstrate the scanning results for $m_{eX^e}$ versus $m_{\mu X^\mu}$ in Fig.~\ref{fig:meXe}, where the filled circles arise from the  constraints shown in Eq.~(\ref{eq:v_nu_mass}), and the squares are derived from the $m_{\ell}$ and $\delta a_\ell$ constraints. 
 According to  the results,  the same parameters from the different phenomena can have the common values.

\begin{figure}[phtb]
\begin{center}
\includegraphics[scale=0.5]{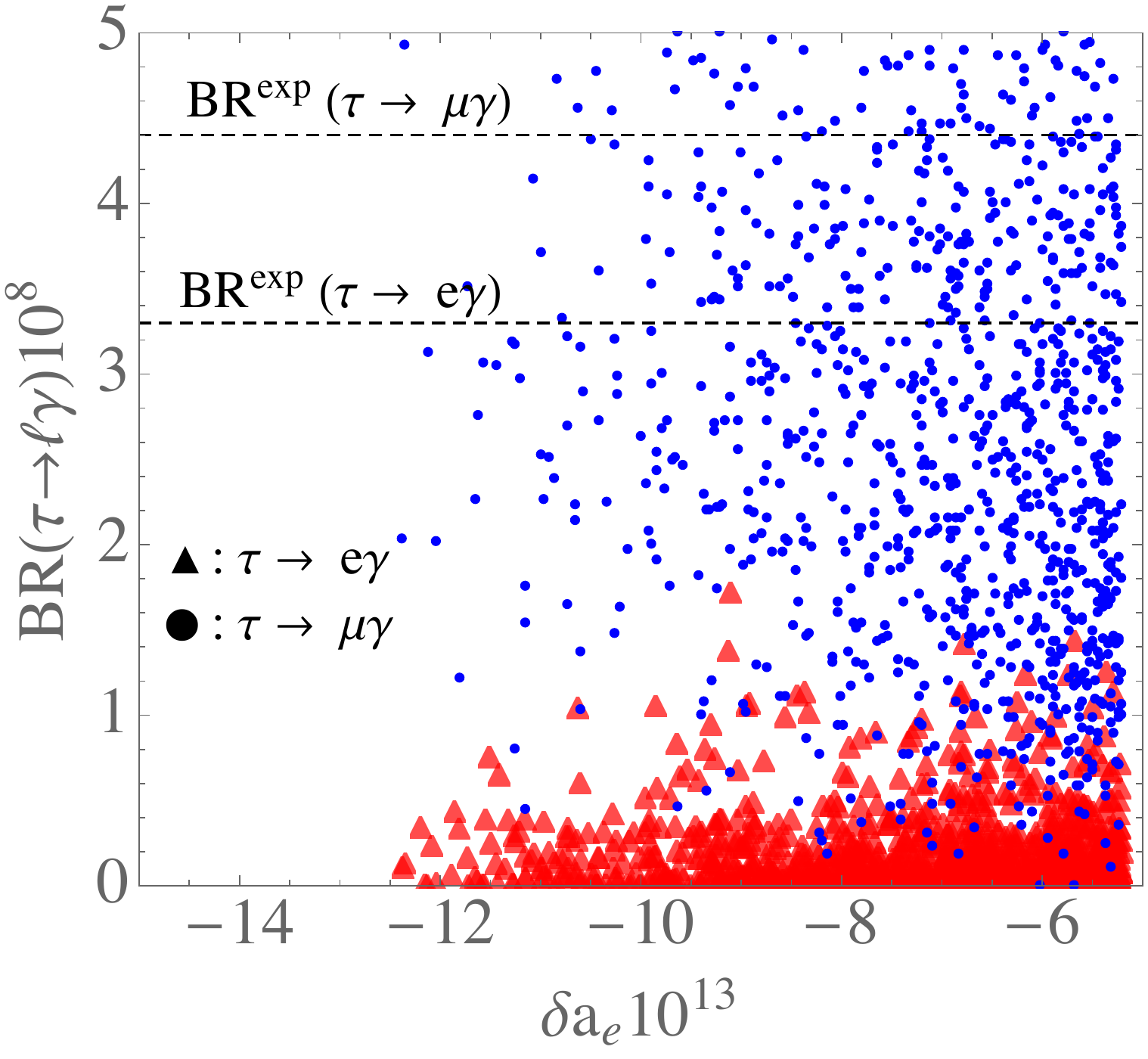}
\includegraphics[scale=0.5]{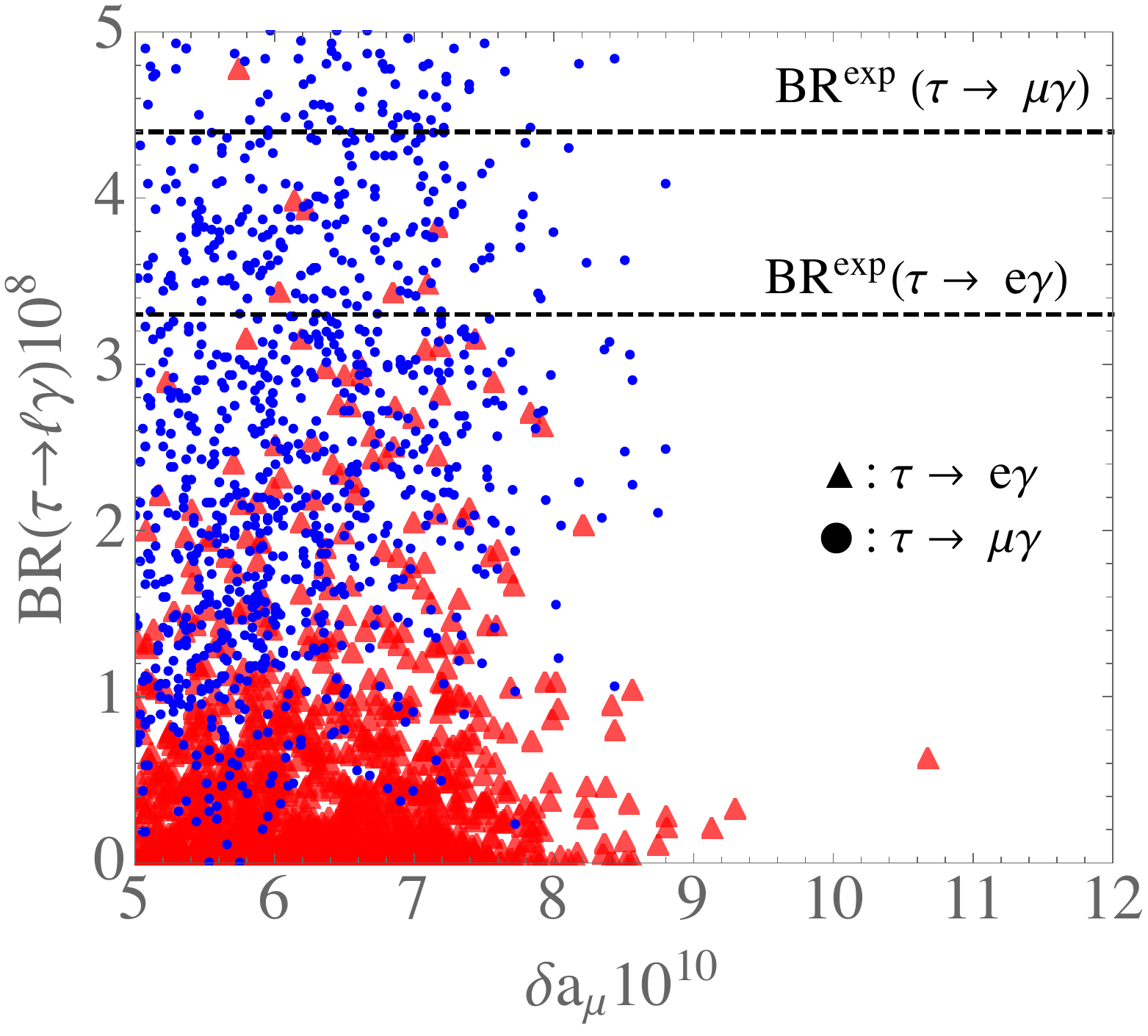}

 \caption{Correlations of  $BR(\tau \to \ell \gamma)$ (in units of $10^{-8}$) with $\delta a_e$ (left panel) and $\delta a_\mu$ (right panel).   }
\label{fig:taulga}
\end{center}
\end{figure}

In the numerical analysis, we used the relations in Eq.~(\ref{eq:mutoe}), where  the rare radiative $\mu \to e \gamma$ decay can be basically as small as the current upper limit.  We thus focus on the situations in the $\tau\to \ell \gamma$ decays.  Using the allowed parameter regions, which are limited by the selected $m_\ell$ and $\delta a_\ell$ regions, 
the BRs in units of $10^{-8}$ for the $\tau \to \ell \gamma$ decays  with respect to $\delta a_e$ are shown in Fig.~\ref{fig:taulga}(a), where the filled circles and triangles denote the $\tau\to e \gamma$ and $\tau \to \mu \gamma$ results, respectively. The correlations of $BR(\tau\to \ell \gamma)$ with $\delta a_\mu$ are given in Fig.~\ref{fig:taulga}(b). From the results, it can be clearly seen that when the upper limits of $BR(\tau\to \ell \gamma)$ are satisfied, $\delta a_e$ of $O(-10^{-12})$ and $\delta a_\mu$ of $O(10^{-9})$ can be achieved. In addition, it is found that with the constrained parameter regions, the resulting $BR(\tau\to \mu \gamma)$ can be over the current upper limit; that is, the $\tau\to \mu \gamma$ decay can further exclude the free parameter space. Nevertheless, when we exclude the sampling points, which are constrained by the $\tau\to \mu \gamma$ decay, the $\delta a_\ell$ results are not changed.

 In addition to the SM-like Higgs boson, the extra scalar bosons, which are directly related to our study, are $s$ and $s^\ell$.  Due to $g_{Z'}\ll 1$, the new scalar couplings to the $Z'$ boson are small.  According to   earlier discussions, the pair production in the $h\to s^\ell s^{\ell'}$ processes can be suppressed by  requiring $m_h < 2 m_{s^\ell}$.  Therefore, from the Yukawa couplings in Eq.~(\ref{eq:Yukawa}), the potential production channels for $s^\ell$ at the LHC can be though the pair production of $X_R^\ell$ and $X_L$, where the associated Yukawa couplings $y^\ell_{\ell'}$ and $y_\ell$ can be of order of unit. For the signal search of $s^\ell$, it depends on the decay channels. From Eq.~(\ref{eq:X_mixing}),  it can be seen that although the  $s^\ell-\ell-\ell$ couplings are small,  the BRs for $s^\ell \to \ell^- \ell^+$ could be $O(1)$. Thus, the favorable signals for probing $s^\ell$ scalars in the $pp$ collisions are via multi-lepton final states:
 \begin{align}
 pp& \to X^\ell_R \bar X^\ell_R \to (\ell_L s^\ell) (\ell^+_L s^\ell) \to (\ell^-_L \ell^+ \ell^-) (\ell^+_L \ell^+ \ell^-)\,, \nonumber \\
  pp& \to X \bar X \to (\ell_R s^\ell) (\ell^+_R s^\ell) \to (\ell_R \ell^+ \ell^-) (\ell^+_R \ell^+ \ell^-)\,.
 \end{align}
 
 Although $S$ doesn't directly couple to the charged leptons, it can couple to the Majorana fermions $N^\ell$. From Eq.~(\ref{eq:s_mixing}), it can be found that the $s-\phi$ mixing, denoted by $U_{s\phi}$, can be $O(0.05)$.  Including the $s-\phi$ mixing, we can obtain the Higgs coupling to $N^{\ell T} N^\ell $ as:
  \begin{equation}
  \frac{1}{\sqrt{2}}  h_\ell U_{s\phi}  N^{\ell T} C N^\ell h\,.
  \end{equation}
 In the study, we take $m_{N^e}=(100,300)$ GeV and $m_{N^\mu}=(10, 30)$ GeV; therefore, the SM Higgs can invisibly decay into $N^\mu N^\mu$, where the partial decay rate can be found as:
  \begin{equation}
  \Gamma(h\to N^\mu N^\mu) = \frac{m_h}{32 \pi} \left| \frac{m_{N^\mu}}{v_S} U_{s \phi}\right|^2 \left( \frac{2 m_{N^\mu}}{m_h}\right)^2 \sqrt{1- \left( \frac{2m_{N^\mu}}{m_h}\right)^2}\,. 
  \end{equation}
  Using $\Gamma^{\rm SM}_h\approx 4.07$ MeV~\cite{PDG}, $U_{s \phi}= -0.085$, and $m_{N^\mu}=30$ GeV, we find that the BR for $h\to N^\mu N^\mu$  is $BR(h\to N^\mu N^\mu) \approx 3.43 \%$,  where the current upper limits at $95 \%$ confidence level  are $26\%$ and $19 \%$ by CMS~\cite{Sirunyan:2018owy} and ATLAS~\cite{Aaboud:2019rtt} experiments,  respectively,  and the SM result is $BR(h \to {\rm invisible}) =1.06 \times 10^{-3} $~\cite{An:2018dwb}.

\section{Summary}

A gauged $U(1)_{e-\mu}$ extension of the SM is used to explain the neutrino masses and the electron and muon $g-2$, where two vector-like lepton doublets, three vector-like lepton singlets, four scalar singlets, and two Majorana fermions are included.  The studying phenomenon are all  generated through the one-loop radiative effects. 

Although the electron and muon do not obtain their masses via the Higgs mechanism, their masses can be induced through  mixing with the introduced heavy charged leptons. We found that the mass hierarchy between the electron and the muon can be accommodated in the model. When the bounds of the electron and muon masses and  the neutrino data are satisfied, we found that the electron $g-2$ can reach an order of $-10^{-12}$, and  the muon $g-2$ can be of an $O(10^{-9})$. 

The radiative lepton-flavor violation processes can arise from similar Feynman diagrams, which are used for producing the lepton $g-2$. When the $\mu\to e \gamma$ decay is suppressed, and the constrained parameter values are applied, the result of $BR(\tau\to e \gamma) \lesssim 10^{-8}$ can be obtained.  With the same constrained parameter set, we found that the resulting $BR(\tau \to \mu \gamma)$ can be larger than its current upper limit; that is, the $\tau \to \mu \gamma$ decay indeed can be used to further constrain the parameter space. Nevertheless, the  parameter regions excluded by the $\tau \to \mu \gamma$ decay do not change  the  regions allowed for  $\delta a_e$ and $\delta a_\mu$. 

 Using the mixing between  the SM Higgs and the new scalar $S$, we found that a significant Higgs invisible decay can be induced in the model, and the associated $BR(h\to {\rm invisible})$ can reach a few percent level.

\section*{Acknowledgments}

 We would like to thank Dr.~Di Liu for useful comments.
 The work was supported in part by KIAS Individual Grants, Grant No. PG054702 (TN) at Korea Institute for Advanced Study.
This work was also supported by the Ministry of Science and Technology of Taiwan,  
under grants MOST-108-2112-M-006-003-MY2.

\end{document}